\documentclass[english]{article}
\usepackage[latin9]{inputenc}
\usepackage{amssymb}
\usepackage{graphicx}

\usepackage{geometry}

\makeatletter

\providecommand{\tabularnewline}{\\}

\providecommand{\tabularnewline}{\\}
\usepackage{babel}

\usepackage{babel}

\usepackage{babel}

\usepackage{babel}

\makeatother

\usepackage{babel}
\begin{document}

\title{The Spin and Orbital Contributions to Magnetic Dipole Transitions}

\author{Arun Kingan, Michael Quinonez , Xiaofei Yu and Larry Zamick\\
 Department of Physics and Astronomy\\
 Rutgers University, Piscataway, New Jersey 08854 }
\maketitle
\begin{abstract}
In previous works we examined the systematics of magnetic dipole
transitions in a single j shell. We here extend the study to large
space calculations. We consider the nuclei $^{44}$Ti, $^{46}$Ti
and $^{48}$Cr. Of particular interest is the contributions to B(M1)
of the spin and orbital parts of the magnetic dipole operator. Whereas
usual scissors mode analyses have as an initial state the J=0+ ground
state (of an even-even nucleus) we here start with the lowest J=1+
T=1 state. This enables us to reach many more states e.g. J=2, T=0, 1,
and 2 and thus get a better picture of the collectivity of this state. We find that the B(M1) strengrh decreases exponentially with energy.  
 \\
 PACs Number: 21.60.Cs 
\end{abstract}

\section{Introduction}

Previously studies of magnetic dipole {[}1,2{]} and Gamow Teller transitions
{[}3{]} in the f$_{7/2}$ shell were performed. The nuclei considered
were $^{44}$Ti and $^{46}$Ti and indeed the model space was the
single j (f$_{7/2}$) shell. In usual scissors mode analyses for even-even
nuclei one starts with a J=0$^{+}$ ground state and the final state
is J=1$^{+}$. In ref {[}2{]}, however, we start with J=1$^{+}$.From
here there are many more places to go, e.g. to J=2$^{+}$ T=0,1,2 .One
thus gets a better picture of the nature of the collectivity of this
state. One striking observation was the fact that there were many strong
B(M1) transitions besides the transition back to the ground state.

In this work we add $^{48}$Cr to the list and most important we extend
the calculation to the full fp shell. Will we also get strong transitions
here? In the single j shell the ratio of the spin and orbit parts
of the B(M1) is unique, i.e. independent of the details of the (single
j) wave functions. However, this is not the case in the large space.
So, we will give the spin, orbit and total B(M1)'s in various tables.

In all table's B(M1) transitions are given in units of $(Mu[suB]N)^{2}$

\section{Tables of B(M1)'s}

In tables I and II we show selected results for  $^{44}$Ti and  $^{46}$Ti of
single j shell calculations of Harper and Zamick {[}1{]}. Mainly, we
show B(M1)'s from the lowest J=1+ T=1 state to I=0+ and I=2+ states
whose values are greater than 0.5. But we also show results to states
of higher isospin even if this is not the case. And we show the transition
to the lowest J=2 T=1 state in  $^{46}$Ti even though the value is 0.246.
This is to emphasize that lowest to lowest are not always large. In
Tables III, IV, and V we show results in the large space for the GX1A interaction
for B(M1)'s greater than 0.05 for  $^{44}$Ti and  $^{48}$Cr, and greater than 0.1 for $^{44}$Ti.

\section{COMMENTS ON TABLES I,II,III IV AND V.}

Table I refers to the previous work of Harper and Zamick{[}1{]}. The
scissors mode is usually identified as the transition from the J=0
ground state to the lowest. J=1 state. In $^{44}$Ti this would be
3 times 1.182=3.546 However if we take J=1+ as our initial state we
see even stronger B(M1)'s which leads us to question what is meant
by collectivity. There is an even stronger B(M1) from J=1 T=1 to J=0 T=2
(1.955) than the ``scissors transition'' (1.182). There is a super
strong transition (12.979) to a J=2 T=0 state at 4.957 MeV. 

Let us compare this single j shell calculation with the large space
calculation in Fig 4.The B(M1) to ground is enhanced from 1.182 to
1.503. The B(M1) to the strong J=0 T=2 state is also enhanced form
1.55 to 2.675.However the superstrong B(M1) is reduced from 12.979
to 5.512. Still the same general pattern is followed when we go from
thesingle j space to the large space.

\begin{quote}
In $^{46}$Ti in the large space (Table IV) the B(M1) from the lowest
J=1 T=1 states to the J=0 T=1 ground state is 0.560 .in the single
j space (Table II) the B(M1) is almost the same as in the large space
0.504. In the small space there is a much larger B(M1) to a (0,1)
state at 4.625 MeV. The value is 2.474. Also, at 6.235 MeV a B(M1)
of 0.675. In the large space there is fragmentation with B(M1)'s to
(0,1) states at 4.294, 5.287, 7.322, 8.167, 8.560 and 10.164 MeV with
respective values of 0.110, 0.927, 0.166, 0.174, 0.422, 0.201 and
0.105. Note that unlike $^{44}$Ti there is here no large B(M1) to
a J=0 T=2 state. This is simply due to the fact that in the single
shell there is no J=0 T=2 state. The only higher isospin state with
J=0 has T=3.One can see this from the fact that $^{46}$Ca is a 2
hole state.

In Table V results are presented for $^{48}$Cr in the large space.
Whereas the B(M1) from the lowest (1,1) state to ground has a value
of 1.101 there are several comparable ones including one to the (0,2)
state at 10.368 MeV with a value of 1.142 and the (2,0) state at 7.206
MeV with a value of 1.104. There is considerable (2,2) fragmentation.
\end{quote}

\section{COMMENTS ON TABLES VI TO XIII}

In tables VI, VII and VII we consider transitions from the lowest (1,1)
state to lowest (J,T) states. The reverse transition (0, T$_{min}$)
to (1,1) is usually associated with the collective scissors mode
state {[}9-25{]}. The review article of K, Heyde, P.von Neumann-Cosel
and A. Richter{[}8{]}contains many more references then what are included
here. The single j results in Tables I and II are closely related to
the single j shell results of Zamick {[}15{]}. 

The B(M1)'s to ground in Table V for the 3 nuclei -44,46 and 48- in
a large space calculation with the GX1A interaction are respectively
1.503,0.504 and 1.101. What is more illuminating is the breakdown
into B(M1)$_{spin}$ and B(M1)$_{0rbit}$. Note that these do not
add up to B(M1)$_{total}$. We have to add amplitudes and then square.

Although, as shown in Table VI the B(M1) in $^{48}$Cr is a bit smaller
than in $^{44}$Ti the B(M1)$_{orbit}$ is considerably larger (0.305
vs. 0.190). This is consistent with the extreme collective picture
of the scissors mode as a pure orbital vibration with a B(M1) proportional
to the deformation parameter. The B(E2) for the lowest 2+ state to
ground is about twice as large in $^{48}$ Cr than in $^{44}$Ti. This
indicates that $^{48}$Cr is more deformed than $^{44}$Ti .

In Table VII we have transitions from J=1 to J=1 for $^{46}$Ti and
$^{48}$Cr. Note the very small results for $^{48}$Cr. This can be
explained by the fact that $^{48}$Cr is at midshell in the single
j shell . One can regard $^{48}$Cr as 8 particles or 8 holes. This
leads to a selection rule for the quantity s= (-1)$^{((Vn+Vp)/2)}$.
In the single j shell the value of s must be the same in the initial
and final state in order to get a non-zero B(M1).

In Table VIII we have for $^{44}$Ti an example of destructive interference
between spin and orbit. The spin value is 0.2449 but the total is 0.1209.
However for $^{48}$Cr the interference is constructive. There is
also destructive interference for $^{44}$Ti in Tables IX and X.

We also consider the decomposition into spin and orbit for cases where the
B(M1)'s are large. These are shown in Tables XI, XII, and XIII. For these cases
the spin and orbit amplitudes add coherently

\section{COMMENTS ON TABLES XIII to XXXI}

We give more detailed tables of B(M1)'s in tables 14 and beyond. We show
results for total, spin and orbit in separate tables. Note than in 48Cr the lowest
J=1+ state has isospin T=0. In the previous tables we used the next state,
the first J=1+ T=1 state, as our starting point. As shown the lowest J=1+
state with T=0 has very weak B(M1)'s to T=0 final states. This is due to the
well known result that the isoscalar M1 coupling is much smaller than the
isovector one. Also in the single j shell limit these T=0 to T=0 B(M1)'s would
vanish [20].

\begin{table}[h]
\global\long\def\thetable{I}
 \centering \caption{$^{44}$Ti Selected B(M1)'s from the lowest 1$^{+}$,T=1 state (E$^{*}=$5.669
MeV) to various states,single j, MBZE int.}
\begin{tabular}{c c c}
\hline 
(I,T)  & E{*} (MeV)  & B(M1) \tabularnewline
\hline 
(0,0)  & 0  & 1.182 \tabularnewline
(0,2)  & 8.284  & 1.955 \tabularnewline
(2,0))  & 1.163  & 1.347 \tabularnewline
(2,0)  & 4.957  & 12.979 \tabularnewline
(2,0)  & 7.823  & 1.097 \tabularnewline
(2,0)  & 7.970  & 1.539 \tabularnewline
(2,1)  & 9.268  & 0 \tabularnewline
(2,2)  & 9.870  & 0.098 \tabularnewline
(2,2)  & 11.822  & 0.114 \tabularnewline
\hline 
\end{tabular}
\end{table}

\begin{table}[h]
\global\long\def\thetable{II}
 \centering \caption{$^{46}$Ti Selected B(M1)'s from lowest 1+ T=1 state(E{*}=3.655 MeV)
to various states, single j, MBZE int.}
\begin{tabular}{c c c}
\hline 
(I,T)  & E{*} (MeV)  & B(M1) \tabularnewline
\hline 
(0,1)  & 0  & 0.560 \tabularnewline
(0,1)  & 4.625  & 2.474 \tabularnewline
(0,1)  & 6.273  & 0.675 \tabularnewline
(2,1)  & 1.148  & 0.203 \tabularnewline
(2,1)  & 2.496  & 1.202 \tabularnewline
(2,1)  & 3.422  & 1.734 \tabularnewline
(2,2)  & 4.883  & 0.246 \tabularnewline
(2,1)  & 5.152  & 0.509 \tabularnewline
(2,1)  & 6.158  & 0.648 \tabularnewline
(2,2)  & 8.255  & 1.622 \tabularnewline
(2,2)  & 9.502  & 0.394 \tabularnewline
(2,2)  & 10.403  & 0.053 \tabularnewline
(2,2)  & 11.898  & 0.005 \tabularnewline
(2,3)  & 14.789  & 0 \tabularnewline
\hline 
\end{tabular}
\end{table}

\begin{table}[h]
\global\long\def\thetable{III}
 \centering \caption{$^{44}$Ti Selected B(M1)'s from the lowest 1$^{+}$T=1 state (E$^{*}$=5.965 MeV) to various states, LARGE SPACE, GX1A }
\begin{tabular}{c c c}
\hline 
(I,T)  & E{*} (MeV)  & B(M1) \tabularnewline
\hline 
(0,0)  & 0  & 1.503 \tabularnewline
(0,0)  & 4.501 & 0.205 \tabularnewline
(0,0)  & 7.634  & 0.098 \tabularnewline
(0,2)  & 8.893 & 2.675 \tabularnewline
(0,0)  & 9.230 & 0.455 \tabularnewline
(0,0)  & 10.189 & 0.233 \tabularnewline
(1,1)  & 5.965  & 0.158 \tabularnewline
(1,0)  & 8.835  & 0.637 \tabularnewline
(1,0)  & 9.572  & 0.057 \tabularnewline
(1,0)  & 10.351  & 0.207 \tabularnewline
(1,0)  & 11.550  & 0.144 \tabularnewline
(1,0)  & 12.186  & 0.136 \tabularnewline
(1,2)  & 17.876  & 0.1022 \tabularnewline
(1,2)  & 18.082  & 0.409 \tabularnewline
(2,0)  & 1.287 & 0.120 \tabularnewline
(2,0)  & 3.168  & 2.013 \tabularnewline
(2,0)  & 6.493 & 0.293 \tabularnewline
(2,0)  & 6.768 & 5.512 \tabularnewline
(2,0)  & 9.659 & 0.193 \tabularnewline
(2,0)  & 10.735 & 0.413 \tabularnewline
\hline 
\end{tabular}
\end{table}

\begin{table}
\global\long\def\thetable{IV}
 \centering \caption{$^{46}$Ti Selected B(M1)'s from the lowest 1$^{+}$T=1state (E$^{*}$=
3.995 MeV) to various states, LARGE SPACE,GX1A}
\begin{tabular}{c c c}
\hline 
(J,T)  & $E$$^{*}$(MeV)  & B(M1) \tabularnewline
\hline 
(0,1)  & 0  & 0.504 \tabularnewline
(0,1)  & 4.294  & 0.110 \tabularnewline
(0,1)  & 5.287  & 0.927 \tabularnewline
(0,1)  & 5.668  & 0.166 \tabularnewline
(0,1)  & 7.322  & 0.174 \tabularnewline
(0,1)  & 8.167  & 0.422 \tabularnewline
(0,1)  & 8.560  & 0.201 \tabularnewline
(0,1)  & 10.164  & 0.105 \tabularnewline
(0,2)  & 15.004  & 0.1527 \tabularnewline
(0,2)  & 15.556  & 0.1604 \tabularnewline
(2,1)  & 1.005  & 0.137 \tabularnewline
(2,1)  & 2.587  & 1.001 \tabularnewline
(2,1)  & 4.281  & 0.408 \tabularnewline
(2,1)  & 5.444  & 0.571 \tabularnewline
(2,1)  & 5.536  & 0.104 \tabularnewline
(2,1)  & 5.815  & 0.122 \tabularnewline
(2,1)  & 6.104  & 0.257 \tabularnewline
(2,1)  & 6.530  & 0.262 \tabularnewline
(2,1)  & 7.675  & .115 \tabularnewline
(2,1)  & 7.994  & 0.398 \tabularnewline
(2,2)  & 8.864  & 1.246 \tabularnewline
(2,2)  & 11.873  & 0.162 \tabularnewline
\hline 
\end{tabular}
\end{table}

\begin{table}[h]
\global\long\def\thetable{V}
 \centering \caption{$^{48}$Cr Selected B(M1)'s from lowest 1+ T=1 state(E{*}=5.5639 MeV)to
various states,LARGE SPACE, GX1A}
\begin{tabular}{c c c}
\hline 
(I,T)  & E{*} (MeV)  & B(M1) \tabularnewline
\hline 
(0,0)  & 0  & 1.101 \tabularnewline
(0,0)  & 5.440  & 0.059 \tabularnewline
(0,0)  & 6.117  & 0.061 \tabularnewline
(0,2)  & 8.571  & 1.142 \tabularnewline
(0,0)  & 9.878  & 0.102 \tabularnewline
(0,0)  & 10.326  & 0.061 \tabularnewline
(2,0)  & 0.788  & 0.466 \tabularnewline
(2,0)  & 3.394  & 0.053 \tabularnewline
(2,0)  & 5.976  & 0.794 \tabularnewline
(2,0)  & 6.412  & 1.090 \tabularnewline
(2,0)  & 6.790  & 0.150 \tabularnewline
(2,0)  & 7.141  & 0.8073 \tabularnewline
(2,0)  & 7.206  & 1.014 \tabularnewline
(2,2)  & 9.582  & 0.160 \tabularnewline
(2,2)  & 12.600  & 0.271 \tabularnewline
(2,2)  & 14.2445  & 0.096 \tabularnewline
(2,2)  & 14.352  & 0.071 \tabularnewline
(2,2)  & 14.711  & 0.236 \tabularnewline
(2,2)  & 14.799  & 0.427 \tabularnewline
(2,2)  & 15.259  & 0.060 \tabularnewline
(2,2)  & 17.2026  & 0.062 \tabularnewline
(2,2)  & 17.454  & 0068 \tabularnewline
(2,2)  & 17.6551  & 0.066 \tabularnewline
(2,2)  & 18.1059  & 0.093 \tabularnewline
\hline 
\end{tabular}
\end{table}

\begin{table}[h]
\global\long\def\thetable{VI}
 \centering \caption{Lowest J=1,T=1 to Ground}
\begin{tabular}{c c c c}
\hline 
 & 44Ti  & 46Ti  & 48Cr \tabularnewline
\hline 
spin  & 0.6249  & 0.1375  & 0.2475 \tabularnewline
oribit  & 0.1897  & 0.1149  & 0.3046 \tabularnewline
total  & 1.503  & 0.5038  & 1.101 \tabularnewline
\hline 
\end{tabular}
\end{table}

\begin{table}[h]
\global\long\def\thetable{VII}
 \centering \caption{Lowest J=1,T=1 to Lowest J=1 T=0}
\begin{tabular}{c c c}
\hline 
 & 44Ti  & 48Cr \tabularnewline
\hline 
spin  & 0.5041  & 0.004357 \tabularnewline
oribit  & 0.007732  & 0.002055 \tabularnewline
total  & 0.6367  & 0.0124 \tabularnewline
\hline 
\end{tabular}
\end{table}

\begin{table}[h]
\global\long\def\thetable{VIII}
 \centering \caption{Lowest J=1,T=1 to Lowest J=2 T=0}
\begin{tabular}{c c c}
\hline 
 & 44Ti  & 48Cr \tabularnewline
\hline 
spin  & 0.2441  & 0.06805 \tabularnewline
oribit  & 0.02141  & 0.1782 \tabularnewline
total  & 0.1209  & 0.4665 \tabularnewline
\hline 
\end{tabular}
\end{table}

\begin{table}[h]
\global\long\def\thetable{IX}
 \centering \caption{Lowest J=1,T=1 to Lowest J=2 T=1}
\begin{tabular}{c c c c}
\hline 
 & 44Ti  & 46Ti  & 48Cr \tabularnewline
\hline 
spin  & 0.02047  & 0.03507  & 0.00002647 \tabularnewline
oribit  & 0.006605  & 0.03343  & 0.000008506 \tabularnewline
total  & 0.003818  & 0.137  & 0.000004964 \tabularnewline
\hline 
\end{tabular}
\end{table}

\begin{table}[h]
\global\long\def\thetable{X}
 \centering \caption{Lowest J=1,T=1 to Lowest J=2 T=2}
\begin{tabular}{c c c c}
\hline 
 & 44Ti  & 46Ti  & 48Cr \tabularnewline
\hline 
spin  & 0.0328 & 0.3905 & 0.01607 \tabularnewline
oribit  & 0.0158  & 0.2416  & 0.0744 \tabularnewline
total  & 0.0031 & 1.246 & 0.1596 \tabularnewline
\hline 
\end{tabular}
\end{table}

\begin{table}[h]
\global\long\def\thetable{XI}
 \centering \caption{44Ti Lowest J=1 T=1 to Select J=0, J=2}
\begin{tabular}{c | c c}
\hline 
(I,T) & (0,2) & (2,0) \tabularnewline
\hline 
EShift & 8.893 & 6.768 \tabularnewline
\hline 
Spin & 1.036 & 2.849 \tabularnewline
Orbit & 0.382 & 0.435 \tabularnewline
Total & 2.675 & 5.512 \tabularnewline
\hline 
\end{tabular}
\end{table}

\begin{table}[h]
\global\long\def\thetable{XII}
 \centering \caption{46Ti Lowest J=1 T=1 to Select J=0, J=2}
\begin{tabular}{c | c c c}
\hline 
(I,T) & (0,1) & (2,1) & (2,1) \tabularnewline
\hline 
EShift & 5.287 & 2.587 & 5.444 \tabularnewline
\hline 
Spin & 0.215 & 0.306 & 0.264 \tabularnewline
Orbit & 0.249 & 0.205 & 0.058 \tabularnewline
Total & 0.927 & 1.011 & 0.571 \tabularnewline
\hline 
\end{tabular}
\end{table}

\begin{table}[h]
\global\long\def\thetable{XIII}
 \centering \caption{48Cr Lowest J=1 T=1 to Select J=0, J=2}
\begin{tabular}{c | c c c}
\hline 
(I,T) & (0,2) & (2,0) & (2,0) \tabularnewline
\hline 
EShift & 8.571 & 6.412 & 7.206 \tabularnewline
\hline 
Spin & 0.263 & 0.450 & 0.344 \tabularnewline
Orbit & 0.308 & 0.140 & 0.177 \tabularnewline
Total & 1.142 & 1.09 & 1.014 \tabularnewline
\hline 
\end{tabular}
\end{table}

\clearpage{}\newpage{}

\begin{figure}
  \includegraphics[width=\linewidth]{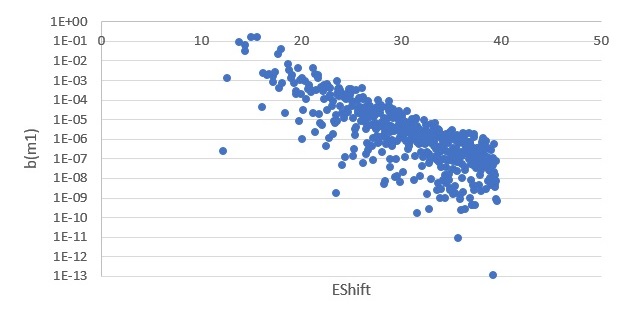}
  \caption{$^{46}$Ti B(M1)'s from Lowest J=1 T=1 to Lowest 500 J=0 T=2, Log Scale }
  \label{fig:46Ti10}
\end{figure}

\begin{figure}
  \includegraphics[width=\linewidth]{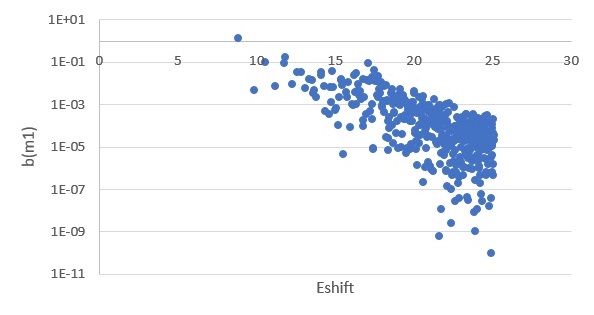}
  \caption{$^{46}$Ti B(M1)'s from Lowest J=1 T=1 to Lowest 430 J=2 T=2, Log Scale }
  \label{fig:46Ti12}
\end{figure}

\begin{figure}
  \includegraphics[width=\linewidth]{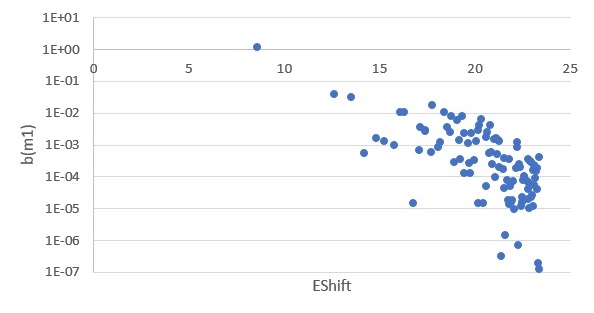}
  \caption{$^{48}$Cr B(M1)'s from Lowest J=1 T=1 to Lowest 100 J=0 T=2, Log Scale}
  \label{fig:48Cr10}
\end{figure}

\begin{figure}
  \includegraphics[width=\linewidth]{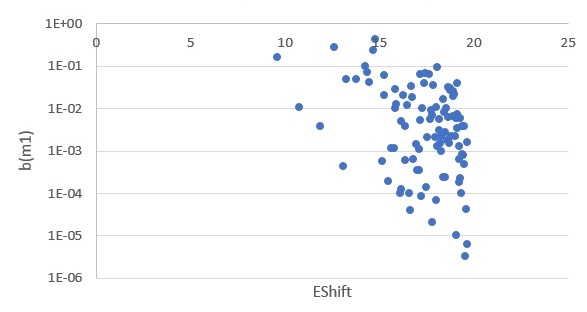}
  \caption{$^{48}$Cr B(M1)'s from Lowest J=1 T=1 to Lowest 100 J=2 T=2, Log Scale}
  \label{fig:48Cr12}
\end{figure}

\clearpage{}\newpage{}

\section{Comments on the Figures}

 In Figues 1 2,3 and 4 we show  a log plot of the B(M1) stengths vs the  energy shift from the
 J=1 T=1 lowest state. In Fig 1 the final sates in 46Ti have quantum numbers 
J=0 T=2  and in Fig2 J=2 T=2. In Figs 3 and 4 we  do the same for 48Cr. .  
In Fig 1 we included 500 states while in Fig 2 430 states. Fo 48Cr we used 
only 100 states and  (Figs III and IV). This is not enough to reach the 
asymptotic region. We will do improved calculations in the near future.
 Although the points  seem at first glance to be all over the place, a closer look shows an exponential 
decrease  of B(M1) strength with energy. one can get a rough fit to Fig1 with
log(B(M1)) = -0.69897 -(E-15) *0.20264            or equivilantly 
B(M1) =0.2 * 10**(-(E-15)*0.20264)   for E greater than 15 MeV.
The important point to make at present is not the precise fit to the calculated
 points but rather to make the simple statement that the strengh , 
displays  an exponential decrease with energy.

\section{Closing remarks}

From the viewpoint of the lowest J=1+ states in $^{44}$Ti, $^{46}$Ti and $^{48}$Cr as initial states,
there are many strong B(M1)'s besides those to the ground state the latter being the
inverse of what is loosely called a scissors or spin scissors transition. In comparing the
full f-p calculation with single-j we see of course more fragmentation in the former and
individual B(M1)'s are not as large. For example, for $^{44}$Ti in single-j there is a very large
transition from J=1+ T=1 at 5.699 MeV to the (2,0) state at 4.954 MeV with B(M1)=12.979.
In the full fp case, Table III, we see (2,0) states at 3.169meV and 6.768 meV with
B(M1)'s of 2.013 and 5.512 respectively. This is a general pattern for other nuclei and
other channels.

We have made a detailed analysis of the spin and orbit contributions to B(M1) and
found that in general they are both important. In earlier works [ ] in more deformed
nuclei such as $^{156}$Gd the picture presented for the excitation from the J=0 ground
state to the J=1+ state of an even-even nucleus was that of an orbital excitation. It
would be of interest to reexamine this region more critically to see how much the spin
contributes. We see some indication in this work that spin goes down with deformation
In comparing $^{44}$Ti and $^{48}$Cr in Table VI we see that in the former B(M1) spin=0.6249
and B(M1) orbit=0.1897. In the more deformed $^{48}$Cr the corresponding values are
0.2475 and 0.3046.

There are also some specific points of interest. We have addressed the issue of
constructive and destructive interference between spin and orbit amplitudes. In Table VII
we see that B(M1)'s from J=1 to J=1 in $^{48}$Cr are much smaller than in $^{44}$Ti. There is a
selection rule in single-j that s = (-1) ((Vp+Vn)/2) should not change. Evidently this
selection rule almost holds even in the large space. Thus, despite the fact that collective 
aspects of B(M1)'s is a much studied subject there is still a lot to be learned.

\section{Acknowledgements}

A.K. received support from the Rutgers Aresty Summer 2016 program
and the Richard J. Plano Research Award Summer 2017. M.Q. has two
institutional affiliations: Rutgers University and Central Connecticut
State University (CCSU). He received support via the Research Experience
for Undergraduates Summer program (REU) from the U.S. National Science
program Grant Number 1263280.
.

\newgeometry{total={184.6mm,254mm}}
\clearpage{}\newpage{}

\begin{table}
\section{Apendix}
\global\long\def\thetable{XIV}
 \centering \caption{$^{48}$Cr J=1 (columns) to J=0 (rows) - Total}
\begin{tabular}{c | c | c c c c c c c}
\hline 
 & Isospin  & 0  & 1  & 0  & 0  & 1 \tabularnewline
\hline 
        & Isospin & 0          & 1          & 0           & 0           & 1           & 1          & 0           \\
\hline
Isopsin & Eshift  & 4.8143     & 5.5639     & 6.5689      & 7.3052      & 7.493       & 7.5829     & 7.975       \\
\hline
0       & 0       & 0.0002018  & 1.101      & 0.01344     & 0.0005902   & 0.6552      & 0.157      & 0.00005432  \\
0       & 3.9697  & 0.00002124 & 0.01221    & 0.0001339   & 0.000008575 & 0.1813      & 0.2143     & 0.002447    \\
0       & 5.4402  & 0.0000119  & 0.05936    & 0.0007667   & 0.00008546  & 0.02178     & 0.004729   & 0.0002072   \\
0       & 6.1168  & 0.0002155  & 0.0612     & 0.001855    & 0.0000196   & 0.3039      & 0.6416     & 0.000459    \\
0       & 7.2218  & 0.00007452 & 0.03035    & 0.0005067   & 1.519E-07   & 0.01579     & 0.2888     & 0.0001154   \\
0       & 7.4451  & 0.001152   & 0.001947   & 0.000938    & 0.0001308   & 0.3756      & 0.2143     & 0.00001691  \\
0       & 7.9405  & 0.00004937 & 0.02728    & 0.000504    & 0.0000647   & 0.207       & 0.71       & 0.00001466  \\
0       & 8.5228  & 0.0001892  & 0.01671    & 0.0007493   & 0.00004306  & 0.04581     & 0.1774     & 0.0001081   \\
2       & 8.571   & 5.198E-09  & 1.142      & 3.514E-09   & 5.785E-09   & 0.05221     & 0.8171     & 4.873E-10   \\
0       & 8.8837  & 0.0004389  & 0.01013    & 0.00001638  & 0.0001048   & 0.0004183   & 0.009838   & 0.00056     \\
1       & 8.8901  & 0.1059     & 0.00004997 & 0.1914      & 0.0007502   & 0.0003649   & 0.00000106 & 0.01855     \\
1       & 9.044   & 0.003645   & 0.0002492  & 0.09059     & 0.0005786   & 0.0002013   & 0.00001093 & 0.1084      \\
0       & 9.136   & 0.00166    & 0.009014   & 0.002769    & 0.00000067  & 0.1078      & 0.034      & 0.001083    \\
0       & 9.3022  & 0.001082   & 0.009839   & 0.00001572  & 0.000002358 & 0.05247     & 0.008491   & 0.000005265 \\
1       & 9.4744  & 0.0168     & 0.0002826  & 0.02173     & 0.004453    & 0.000006834 & 0.0001241  & 0.1683      \\
0       & 9.6732  & 0.00004878 & 0.02568    & 0.0002304   & 0.00002637  & 0.03994     & 0.001503   & 0.0007528   \\
0       & 9.8784  & 0.0001671  & 0.1021     & 0.0001102   & 8.623E-07   & 0.04465     & 0.1051     & 0.000005801 \\
0       & 10.1087 & 0.0003918  & 0.04832    & 0.000003111 & 0.00002538  & 0.01002     & 0.005876   & 0.0001179   \\
1       & 10.2023 & 0.2357     & 0.00000483 & 0.004278    & 0.1618      & 0.000408    & 0.00002523 & 0.000001875 \\
0       & 10.3264 & 0.00001633 & 0.06066    & 0.0009946   & 0.00002302  & 0.005904    & 0.00007965 & 0.00007559 \\
\hline 
\end{tabular}

\bigskip

\global\long\def\thetable{XV}
\centering \caption{$^{48}$Cr J=1 (columns) to J=2 (rows) - Total}
\begin{tabular}{c | c | c c c c c c c}
\hline 
        & Isospin & 0           & 1           & 0           & 0           & 1           & 1          & 0          \\
\hline
Isopsin & Eshift  & 4.8143      & 5.5639      & 6.5689      & 7.3052      & 7.493       & 7.5829     & 7.975      \\
\hline
0       & 0.7882  & 0.00001095  & 0.4665      & 0.005434    & 0.00005295  & 0.3838      & 0.09775    & 0.00004686 \\
0       & 3.3937  & 0.00006794  & 0.05316     & 0.000128    & 0.00007609  & 0.004567    & 0.6353     & 0.003363   \\
0       & 4.0979  & 0.002204    & 0.005641    & 0.002974    & 0.00008215  & 0.2153      & 0.08153    & 0.00002041 \\
0       & 4.6235  & 0.0001031   & 0.0003792   & 0.00002682  & 0.0001078   & 0.002023    & 0.003323   & 0.0002398  \\
1       & 5.4949  & 0.05158     & 0.000005071 & 0.06302     & 0.02379     & 0.000006831 & 0.0003423  & 0.0005235  \\
0       & 5.6893  & 0.0002817   & 0.04853     & 0.000007818 & 0.0001299   & 0.0001461   & 0.01539    & 0.00002706 \\
0       & 5.9759  & 0.0002496   & 0.7936      & 0.002608    & 2.203E-07   & 0.02606     & 0.02468    & 0.0001703  \\
0       & 6.4115  & 0.0009975   & 1.09        & 0.002143    & 0.0002732   & 0.1381      & 0.1055     & 0.0007254  \\
1       & 6.7728  & 0.03936     & 0.00007882  & 0.01887     & 0.0001221   & 0.000006557 & 0.00006462 & 0.000053   \\
0       & 6.7845  & 0.0005807   & 0.1498      & 0.0009046   & 0.0001527   & 0.2163      & 0.07794    & 0.00008037 \\
0       & 6.8953  & 0.000001485 & 0.007883    & 0.0001161   & 0.000006219 & 0.1359      & 0.5137     & 2.522E-07  \\
1       & 6.9752  & 0.04337     & 0.001817    & 0.01825     & 0.0001085   & 0.003535    & 0.0001594  & 0.2129     \\
0       & 7.1414  & 0.0001512   & 0.8073      & 0.0004223   & 0.0006917   & 0.0002711   & 0.1442     & 0.00008762 \\
0       & 7.2063  & 0.0005591   & 1.014       & 0.000626    & 0.00004292  & 0.02431     & 0.1373     & 0.00001994 \\
0       & 7.4083  & 0.000408    & 0.01839     & 0.00205     & 0.0002064   & 0.0005325   & 0.1457     & 0.0001307  \\
1       & 7.6538  & 0.03139     & 0.000007927 & 0.1081      & 0.06269     & 0.00005276  & 0.0001034  & 0.001731   \\
1       & 7.9707  & 0.02992     & 0.00009321  & 0.003827    & 0.002171    & 0.00003373  & 0.00007677 & 0.639      \\
0       & 8.0024  & 0.00001222  & 0.01146     & 0.000005752 & 0.00001029  & 0.004944    & 0.004218   & 0.00000311 \\
0       & 8.0476  & 0.0004032   & 0.003086    & 0.0000209   & 0.0001625   & 0.2737      & 2.082      & 7.679E-07  \\
1       & 8.1318  & 0.08284     & 0.0001986   & 0.1484      & 0.007042    & 0.001189    & 1.606E-08  & 0.0007479 \\
\hline 
\end{tabular}
\end{table}

\clearpage{}\newpage{}

\begin{table}
\global\long\def\thetable{XVI}
 \centering \caption{$^{48}$Cr J=1 (columns) to J=0 (rows) - Spin}
\begin{tabular}{c | c | c c c c c c c}
\hline 
        & Isospin & 0          & 1           & 0          & 0           & 1          & 1          & 0          \\
\hline
Isopsin & Eshift  & 4.8143     & 5.5639      & 6.5689     & 7.3052      & 7.493      & 7.5829     & 7.975      \\
\hline
0       & 0       & 0.001078   & 0.2475      & 0.07202    & 0.003167    & 0.4735     & 0.1021     & 0.0002912  \\
0       & 3.9697  & 0.0001142  & 0.000005291 & 0.0007147  & 0.00004783  & 0.0412     & 0.03663    & 0.01312    \\
0       & 5.4402  & 0.00006437 & 0.03027     & 0.0041     & 0.0004549   & 0.0007968  & 0.01089    & 0.001111   \\
0       & 6.1168  & 0.001153   & 0.005221    & 0.009956   & 0.0001082   & 0.1447     & 0.1613     & 0.002456   \\
0       & 7.2218  & 0.0003999  & 0.008146    & 0.002724   & 7.832E-07   & 0.0002444  & 0.07024    & 0.000618   \\
0       & 7.4451  & 0.006165   & 0.007671    & 0.00502    & 0.0007048   & 0.196      & 0.04383    & 0.0000905  \\
0       & 7.9405  & 0.000264   & 0.02776     & 0.002696   & 0.0003465   & 0.05543    & 0.2608     & 0.00007911 \\
0       & 8.5228  & 0.00102    & 0.0006417   & 0.004008   & 0.0002318   & 0.06442    & 0.08389    & 0.0005834  \\
2       & 8.571   & 2.985E-10  & 0.2633      & 3.995E-09  & 5.651E-10   & 0.000546   & 0.2748     & 1.038E-09  \\
0       & 8.8837  & 0.001882   & 0.01353     & 0.00001926 & 0.0005875   & 0.002305   & 0.0006984  & 0.003199   \\
1       & 8.8901  & 0.004106   & 0.0002906   & 0.1268     & 0.000002153 & 0.002031   & 7.302E-07  & 0.003534   \\
1       & 9.044   & 0.0102     & 0.001345    & 0.06178    & 0.00003125  & 0.001073   & 0.00005587 & 0.08215    \\
0       & 9.136   & 0.008913   & 0.001112    & 0.01489    & 0.000003871 & 0.1466     & 0.005426   & 0.005829   \\
0       & 9.3022  & 0.005804   & 0.0001281   & 0.00008398 & 0.00001252  & 0.02347    & 0.009417   & 0.00002774 \\
1       & 9.4744  & 0.01136    & 0.001518    & 0.01554    & 0.005445    & 0.00003574 & 0.0006659  & 0.1344     \\
0       & 9.6732  & 0.0002608  & 0.004276    & 0.001234   & 0.0001415   & 0.03534    & 0.003736   & 0.004036   \\
0       & 9.8784  & 0.000895   & 0.05549     & 0.0005899  & 0.000004552 & 0.02393    & 0.05267    & 0.00003196 \\
0       & 10.1087 & 0.002093   & 0.02332     & 0.00001654 & 0.000135    & 0.0009916  & 0.005266   & 0.0006291  \\
1       & 10.2023 & 0.08753    & 0.00002685  & 0.01075    & 0.03976     & 0.002187   & 0.0001373  & 0.009837   \\
0       & 10.3264 & 0.00008101 & 0.06895     & 0.005332   & 0.00013     & 0.0005298  & 0.002683   & 0.0004013 \\
\hline 
\end{tabular}

\bigskip

\global\long\def\thetable{XVII}
 \centering \caption{$^{48}$Cr J=1 (columns) to J=2 (rows) - Spin}
\begin{tabular}{c | c | c c c c c c c}
\hline 
        & Isospin & 0           & 1          & 0          & 0           & 1          & 1         & 0           \\
\hline
Isopsin & Eshift  & 4.8143      & 5.5639     & 6.5689     & 7.3052      & 7.493      & 7.5829    & 7.975       \\
\hline
0       & 0.7882  & 0.00005861  & 0.06805    & 0.02913    & 0.000282    & 0.2284     & 0.0704    & 0.0002498   \\
0       & 3.3937  & 0.000364    & 0.008086   & 0.0006907  & 0.0004077   & 0.002422   & 0.1578    & 0.01802     \\
0       & 4.0979  & 0.01181     & 0.01365    & 0.01594    & 0.0004398   & 0.1878     & 0.06674   & 0.000111    \\
0       & 4.6235  & 0.0005517   & 0.00003518 & 0.0001448  & 0.000578    & 0.00118    & 0.01394   & 0.001288    \\
1       & 5.4949  & 0.03448     & 0.00002677 & 0.07078    & 0.02995     & 0.00003674 & 0.001838  & 0.003087    \\
0       & 5.6893  & 0.001509    & 0.002483   & 0.00004214 & 0.0006958   & 0.01282    & 0.0004831 & 0.0001467   \\
0       & 5.9759  & 0.001336    & 0.07577    & 0.01399    & 0.000001446 & 0.006259   & 0.0246    & 0.0009103   \\
0       & 6.4115  & 0.005354    & 0.4497     & 0.0115     & 0.00146     & 0.09451    & 0.04323   & 0.003896    \\
1       & 6.7728  & 0.01913     & 0.0002779  & 0.01621    & 0.0001822   & 0.00008292 & 0.0004391 & 0.005878    \\
0       & 6.7845  & 0.002933    & 0.03232    & 0.004992   & 0.0008089   & 0.1616     & 0.0136    & 0.0004457   \\
0       & 6.8953  & 0.000007583 & 0.0001863  & 0.0006273  & 0.00003193  & 0.0376     & 0.211     & 0.000001709 \\
1       & 6.9752  & 0.2883      & 0.009697   & 0.03167    & 0.003013    & 0.01892    & 0.0008694 & 0.09339     \\
0       & 7.1414  & 0.0008128   & 0.1938     & 0.002271   & 0.003715    & 0.0002151  & 0.1065    & 0.0004669   \\
0       & 7.2063  & 0.002998    & 0.3441     & 0.003353   & 0.0002289   & 0.01424    & 0.05547   & 0.0001072   \\
0       & 7.4083  & 0.002184    & 0.007546   & 0.01099    & 0.001105    & 0.009122   & 0.07594   & 0.0007016   \\
1       & 7.6538  & 0.003735    & 0.00004219 & 0.131      & 0.01208     & 0.0002864  & 0.0005571 & 0.001413    \\
1       & 7.9707  & 0.005715    & 0.0004706  & 0.01933    & 0.0001243   & 0.00016    & 0.0003782 & 0.2329      \\
0       & 8.0024  & 0.00008637  & 0.004253   & 0.00003027 & 0.00004895  & 0.003858   & 0.01354   & 8.748E-07   \\
0       & 8.0476  & 0.002175    & 0.0008109  & 0.0001114  & 0.0008624   & 0.0721     & 0.593     & 0.000002696 \\
1       & 8.1318  & 0.01577     & 0.001068   & 0.2711     & 0.001164    & 0.006406   & 3.502E-08 & 0.003194   \\
\hline 
\end{tabular}
\end{table}

\clearpage{}\newpage{}

\begin{table}
\global\long\def\thetable{XVIII}
 \centering \caption{$^{48}$Cr J=1 (columns) to J=0 (rows) - Orbit}
\begin{tabular}{c | c | c c c c c c c}
\hline 
        & Isospin & 0          & 1           & 0           & 0           & 1          & 1          & 0           \\
\hline
Isopsin & Eshift  & 4.8143     & 5.5639      & 6.5689      & 7.3052      & 7.493      & 7.5829     & 7.975       \\
\hline
0       & 0       & 0.0003471  & 0.3046      & 0.02324     & 0.001023    & 0.01472    & 0.005894   & 0.00009396  \\
0       & 3.9697  & 0.00003697 & 0.0117      & 0.0002298   & 0.0000159   & 0.04965    & 0.07371    & 0.004234    \\
0       & 5.4402  & 0.00002092 & 0.004853    & 0.00132     & 0.000146    & 0.03091    & 0.001267   & 0.0003587   \\
0       & 6.1168  & 0.0003719  & 0.03067     & 0.003217    & 0.00003567  & 0.02921    & 0.1595     & 0.0007914   \\
0       & 7.2218  & 0.0001292  & 0.007051    & 0.0008809   & 2.452E-07   & 0.01996    & 0.07418    & 0.0001993   \\
0       & 7.4451  & 0.001987   & 0.01735     & 0.001618    & 0.0002283   & 0.02896    & 0.06427    & 0.00002918  \\
0       & 7.9405  & 0.00008504 & 0.000002109 & 0.0008685   & 0.0001117   & 0.04821    & 0.1102     & 0.00002567  \\
0       & 8.5228  & 0.0003306  & 0.0108      & 0.001292    & 0.00007508  & 0.001582   & 0.01731    & 0.0001893   \\
2       & 8.571   & 3.005E-09  & 0.3083      & 1.544E-11   & 2.734E-09   & 0.04207    & 0.1442     & 1.029E-10   \\
0       & 8.8837  & 0.0005033  & 0.0002452   & 1.168E-07   & 0.000196    & 0.004687   & 0.005294   & 0.001082    \\
1       & 8.8901  & 0.06829    & 0.00009956  & 0.006645    & 0.0008327   & 0.000674   & 3.063E-08  & 0.005891    \\
1       & 9.044   & 0.001648   & 0.0004361   & 0.002749    & 0.0008788   & 0.0003446  & 0.00001738 & 0.001822    \\
0       & 9.136   & 0.00288    & 0.003795    & 0.004815    & 0.00000132  & 0.002988   & 0.01226    & 0.001887    \\
0       & 9.3022  & 0.001874   & 0.01221     & 0.00002704  & 0.000004009 & 0.005757   & 0.00002398 & 0.000008834 \\
1       & 9.4744  & 0.0005315  & 0.0004909   & 0.0005179   & 0.00004983  & 0.00001132 & 0.0002151  & 0.001903    \\
0       & 9.6732  & 0.00008401 & 0.009       & 0.0003982   & 0.00004573  & 0.0001406  & 0.0004998  & 0.001303    \\
0       & 9.8784  & 0.0002886  & 0.007062    & 0.0001901   & 0.000001452 & 0.003208   & 0.008979   & 0.00001053  \\
0       & 10.1087 & 0.0006739  & 0.004503    & 0.000005306 & 0.00004329  & 0.004708   & 0.0000167  & 0.0002023   \\
1       & 10.2023 & 0.03595    & 0.000008906 & 0.001463    & 0.04115     & 0.000706   & 0.00004484 & 0.01011     \\
0       & 10.3264 & 0.00002459 & 0.0002658   & 0.001721    & 0.00004362  & 0.002896   & 0.003688   & 0.0001286  \\
\hline 
\end{tabular}

\bigskip

\global\long\def\thetable{XIX}
 \centering \caption{$^{48}$Cr J=1 (columns) to J=2 (rows) - Orbit}
\begin{tabular}{c | c | c c c c c c c}
\hline 
        & Isospin & 0           & 1           & 0           & 0           & 1           & 1         & 0           \\
\hline
Isopsin & Eshift  & 4.8143      & 5.5639      & 6.5689      & 7.3052      & 7.493       & 7.5829    & 7.975       \\
\hline
0       & 0.7882  & 0.0000189   & 0.1782      & 0.009399    & 0.00009057  & 0.02004     & 0.00224   & 0.00008024  \\
0       & 3.3937  & 0.0001174   & 0.01978     & 0.0002241   & 0.0001315   & 0.0003374   & 0.1598    & 0.005814    \\
0       & 4.0979  & 0.003809    & 0.00174     & 0.005144    & 0.0001418   & 0.0009339   & 0.0007389 & 0.00003622  \\
0       & 4.6235  & 0.0001778   & 0.0001834   & 0.00004697  & 0.0001865   & 0.006294    & 0.003649  & 0.0004164   \\
1       & 5.4949  & 0.001717    & 0.000008536 & 0.0002252   & 0.0003553   & 0.00001189  & 0.0005938 & 0.001068    \\
0       & 5.6893  & 0.0004868   & 0.02906     & 0.00001366  & 0.0002244   & 0.01571     & 0.01042   & 0.00004775  \\
0       & 5.9759  & 0.0004306   & 0.3789      & 0.004518    & 5.372E-07   & 0.006775    & 6.97E-08  & 0.0002931   \\
0       & 6.4115  & 0.001729    & 0.1396      & 0.003712    & 0.0004703   & 0.004119    & 0.01368   & 0.001259    \\
1       & 6.7728  & 0.003611    & 0.00006073  & 0.000101    & 0.0006025   & 0.00004284  & 0.0001668 & 0.004815    \\
0       & 6.7845  & 0.0009038   & 0.04296     & 0.001646    & 0.0002587   & 0.003991    & 0.02642   & 0.0001476   \\
0       & 6.8953  & 0.000002357 & 0.005646    & 0.0002037   & 0.000009965 & 0.03055     & 0.06623   & 6.483E-07   \\
1       & 6.9752  & 0.108       & 0.003118    & 0.001838    & 0.004265    & 0.006099    & 0.0002842 & 0.02428     \\
0       & 7.1414  & 0.0002628   & 0.21        & 0.0007345   & 0.001201    & 0.000003237 & 0.002845  & 0.00015     \\
0       & 7.2063  & 0.0009679   & 0.1766      & 0.001082    & 0.00007359  & 0.001338    & 0.01822   & 0.00003465  \\
0       & 7.4083  & 0.0007042   & 0.002375    & 0.003548    & 0.0003564   & 0.01406     & 0.01127   & 0.0002266   \\
1       & 7.6538  & 0.01347     & 0.00001354  & 0.001098    & 0.01973     & 0.00009332  & 0.0001805 & 0.00001618  \\
1       & 7.9707  & 0.00948     & 0.0001449   & 0.005958    & 0.003335    & 0.00004679  & 0.0001142 & 0.1004      \\
0       & 8.0024  & 0.00003362  & 0.001749    & 0.000009631 & 0.00001435  & 0.00006725  & 0.002645  & 0.000007284 \\
0       & 8.0476  & 0.0007055   & 0.007061    & 0.00003578  & 0.0002762   & 0.06484     & 0.4529    & 5.863E-07   \\
1       & 8.1318  & 0.02633     & 0.0003458   & 0.01834     & 0.01393     & 0.002075    & 3.651E-09 & 0.0008506  \\
\hline 
\end{tabular}
\end{table}

\clearpage{}\newpage{}

\begin{table}
\global\long\def\thetable{XX}
 \centering \caption{$^{46}$Ti J=1 (columns) to J=0 (rows) - Total}
\begin{tabular}{c | c | c c c c c c c}
\hline 
        & Isospin & 1         & 1           & 1           & 1         & 1         & 1         & 1        \\
\hline
Isopsin & Eshift  & 3.9959    & 5.6008      & 6.0668      & 6.6919    & 7.0243    & 7.1855    & 7.4875   \\
\hline
1       & 0       & 0.5038    & 0.3299      & 0.1594      & 0.08954   & 0.002479  & 0.1151    & 0.01073  \\
1       & 4.2488  & 0.1099    & 0.006629    & 0.4388      & 0.01479   & 0.002153  & 0.2601    & 0.09493  \\
1       & 5.2874  & 0.9267    & 0.3252      & 0.004185    & 0.02523   & 0.01148   & 0.07647   & 0.1264   \\
1       & 5.6685  & 0.166     & 0.2187      & 0.05332     & 0.4146    & 0.2696    & 0.03424   & 0.002201 \\
1       & 6.58    & 0.003196  & 0.3026      & 1.026       & 0.5416    & 0.1852    & 0.0355    & 0.01416  \\
1       & 6.8885  & 0.0004284 & 0.001502    & 0.01674     & 0.0005977 & 0.08463   & 0.3573    & 0.131    \\
1       & 7.3222  & 0.1743    & 0.1209      & 0.06612     & 0.0001065 & 0.003777  & 0.002255  & 0.05149  \\
1       & 8.1666  & 0.4224    & 0.0054      & 0.3951      & 0.06067   & 0.001271  & 0.2833    & 0.06379  \\
1       & 8.5605  & 0.2011    & 0.0687      & 0.2321      & 0.1168    & 0.1412    & 0.0175    & 0.1413   \\
1       & 9.0534  & 0.04214   & 0.08437     & 0.01337     & 0.05393   & 0.02135   & 0.001074  & 0.1099   \\
1       & 9.4385  & 0.02446   & 0.07363     & 0.007747    & 0.004629  & 0.004564  & 0.00215   & 0.06312  \\
1       & 9.9011  & 0.0117    & 0.0239      & 0.009005    & 0.1787    & 0.08442   & 0.00163   & 0.1906   \\
1       & 10.0432 & 0.002472  & 0.00918     & 0.000009334 & 0.01837   & 0.009254  & 0.1022    & 0.1702   \\
1       & 10.1636 & 0.1047    & 0.009813    & 0.04784     & 0.03467   & 0.08414   & 0.428     & 0.1444   \\
1       & 10.3307 & 0.00129   & 0.1313      & 0.02198     & 0.04686   & 0.05013   & 0.1457    & 0.07722  \\
1       & 10.6885 & 0.02451   & 0.000006322 & 0.007882    & 0.02349   & 0.0004393 & 0.08701   & 0.1787   \\
1       & 10.8004 & 0.0001509 & 0.00652     & 0.002898    & 0.1293    & 0.03066   & 0.009285  & 0.1873   \\
1       & 11.0128 & 0.05207   & 0.008879    & 0.02266     & 0.01912   & 0.008989  & 0.0205    & 0.03124  \\
1       & 11.2704 & 0.004107  & 0.01102     & 0.02116     & 0.05189   & 0.002901  & 0.07269   & 0.02635  \\
1       & 11.4117 & 0.008155  & 0.08892     & 0.02348     & 0.0003734 & 0.003076  & 0.0002348 & 0.001001 \\
\hline 
\end{tabular}

\bigskip

\global\long\def\thetable{XXI}
 \centering \caption{$^{46}$Ti J=1 (columns) to J=2 (rows) - Total}
\begin{tabular}{c | c | c c c c c c c}
\hline 
        & Isospin & 1          & 1          & 1           & 1        & 1        & 1         & 1        \\
\hline
Isopsin & Eshift  & 3.9959     & 5.6008     & 6.0668      & 6.6919   & 7.0243   & 7.1855    & 7.4875   \\
\hline
1       & 1.0055  & 0.137      & 1.292E-08  & 0.1297      & 0.08034  & 0.1469   & 0.2743    & 0.1548   \\
1       & 2.5869  & 1.011      & 0.4173     & 0.006072    & 0.08213  & 0.2391   & 0.228     & 0.4812   \\
1       & 3.3857  & 0.01606    & 0.1176     & 0.07582     & 0.05703  & 0.05392  & 0.1872    & 0.01263  \\
1       & 4.2806  & 0.4083     & 0.0005119  & 0.7718      & 0.306    & 0.003245 & 0.3848    & 0.004055 \\
1       & 5.0092  & 0.09655    & 0.02561    & 0.000004616 & 0.01141  & 0.06842  & 0.116     & 0.02561  \\
1       & 5.4438  & 0.5707     & 0.02872    & 0.007544    & 0.1228   & 0.1562   & 0.09917   & 0.08886  \\
1       & 5.5356  & 0.1038     & 0.2526     & 0.001184    & 0.0399   & 0.007906 & 0.1208    & 0.3721   \\
1       & 5.8152  & 0.1216     & 0.01334    & 0.3903      & 0.008193 & 0.1118   & 0.01456   & 0.0509   \\
1       & 6.1044  & 0.2527     & 0.1299     & 0.2901      & 0.4732   & 0.009902 & 0.0006186 & 0.06477  \\
1       & 6.377   & 0.008253   & 0.494      & 0.002221    & 0.07783  & 0.03439  & 0.07168   & 0.01206  \\
1       & 6.53    & 0.2615     & 0.0119     & 0.3329      & 0.0446   & 0.04023  & 0.01032   & 0.1417   \\
1       & 6.6492  & 0.02755    & 0.001464   & 0.04125     & 0.04842  & 0.02069  & 0.02098   & 0.005018 \\
1       & 6.9034  & 0.03039    & 0.1081     & 0.04099     & 0.01281  & 0.00845  & 0.3642    & 0.02272  \\
1       & 7.0252  & 0.002331   & 0.01624    & 0.001543    & 0.001447 & 0.01817  & 0.003733  & 0.175    \\
1       & 7.1253  & 0.09014    & 0.00004846 & 0.07025     & 0.0481   & 0.2341   & 0.007492  & 0.1854   \\
1       & 7.4944  & 0.00002165 & 0.002169   & 0.02774     & 0.3187   & 0.0126   & 0.3167    & 0.1537   \\
1       & 7.5839  & 0.002361   & 0.104      & 0.2768      & 0.001614 & 0.04721  & 0.2958    & 0.004094 \\
1       & 7.6748  & 0.1151     & 0.1332     & 4.08E-08    & 0.3083   & 0.008662 & 0.2876    & 0.2314   \\
1       & 7.8701  & 0.0827     & 0.0204     & 0.02315     & 0.004434 & 0.02664  & 0.07715   & 0.3088   \\
1       & 7.9942  & 0.3982     & 0.0007066  & 0.009904    & 0.2867   & 0.05341  & 0.0003499 & 0.1905  \\
\hline 
\end{tabular}
\end{table}

\clearpage{}\newpage{}

\begin{table}
\global\long\def\thetable{XXII}
 \centering \caption{$^{46}$Ti J=1 (columns) to J=0 (rows) - Spin}
\begin{tabular}{c | c | c c c c c c c}
\hline 
        & Isospin & 1        & 1           & 1        & 1         & 1         & 1         & 1         \\
\hline
Isopsin & Eshift  & 3.9959   & 5.6008      & 6.0668   & 6.6919    & 7.0243    & 7.1855    & 7.4875    \\
\hline
1       & 0       & 0.1375   & 0.1441      & 0.07236  & 0.09055   & 0.000762  & 0.1857    & 0.03711   \\
1       & 4.2488  & 0.01207  & 0.008987    & 0.1128   & 0.01465   & 0.02023   & 0.09579   & 0.03225   \\
1       & 5.2874  & 0.2147   & 0.1405      & 0.01024  & 0.02439   & 0.0009053 & 0.05782   & 0.03908   \\
1       & 5.6685  & 0.08161  & 0.06408     & 0.05078  & 0.1265    & 0.06075   & 0.005107  & 0.0007833 \\
1       & 6.58    & 0.002055 & 0.1316      & 0.322    & 0.1972    & 0.05299   & 0.003393  & 0.002128  \\
1       & 6.8885  & 0.001688 & 0.000001636 & 0.002714 & 0.00118   & 0.02631   & 0.2012    & 0.1182    \\
1       & 7.3222  & 0.0634   & 0.04243     & 0.02317  & 0.0002511 & 0.0002175 & 0.002575  & 0.02796   \\
1       & 8.1666  & 0.1141   & 0.0004286   & 0.1387   & 0.03979   & 0.01165   & 0.08304   & 0.003676  \\
1       & 8.5605  & 0.07651  & 0.04627     & 0.09148  & 0.02751   & 0.02479   & 0.02576   & 0.08279   \\
1       & 9.0534  & 0.01806  & 0.03948     & 0.01471  & 0.04827   & 0.007842  & 0.005056  & 0.03433   \\
1       & 9.4385  & 0.03957  & 0.02025     & 0.00154  & 0.00938   & 0.01181   & 0.0002372 & 0.01325   \\
1       & 9.9011  & 0.0165   & 0.005261    & 0.007655 & 0.06135   & 0.01288   & 0.00795   & 0.1044    \\
1       & 10.0432 & 0.002979 & 0.00611     & 0.003313 & 0.01085   & 0.008227  & 0.07304   & 0.154     \\
1       & 10.1636 & 0.09761  & 0.0006785   & 0.03903  & 0.003706  & 0.05793   & 0.2528    & 0.1072    \\
1       & 10.3307 & 0.009873 & 0.07133     & 0.02005  & 0.03653   & 0.02242   & 0.1184    & 0.04045   \\
1       & 10.6885 & 0.008109 & 0.002536    & 0.006986 & 0.02866   & 0.003217  & 0.09571   & 0.2041    \\
1       & 10.8004 & 0.002388 & 0.000793    & 0.003729 & 0.08256   & 0.02926   & 0.01685   & 0.1231    \\
1       & 11.0128 & 0.04523  & 0.02127     & 0.05382  & 0.007789  & 0.007417  & 0.02484   & 0.02033   \\
1       & 11.2704 & 0.002329 & 0.0308      & 0.02313  & 0.05133   & 0.00155   & 0.02521   & 0.03364   \\
1       & 11.4117 & 0.01021  & 0.06054     & 0.01115  & 0.004801  & 0.003651  & 0.001705  & 0.0001408 \\
\hline 
\end{tabular}

\bigskip

\global\long\def\thetable{XXIII}
 \centering \caption{$^{46}$Ti J=1 (columns) to J=2 (rows) - Spin}
\begin{tabular}{c | c | c c c c c c c}
\hline 
        & Isospin & 1         & 1         & 1          & 1           & 1           & 1         & 1        \\
\hline
Isopsin & Eshift  & 3.9959    & 5.6008    & 6.0668     & 6.6919      & 7.0243      & 7.1855    & 7.4875   \\
\hline
1       & 1.0055  & 0.03508   & 0.001898  & 0.01209    & 0.07544     & 0.06427     & 0.2302    & 0.2318   \\
1       & 2.5869  & 0.3058    & 0.1665    & 0.003974   & 0.05238     & 0.08733     & 0.2781    & 0.5442   \\
1       & 3.3857  & 0.01267   & 0.03811   & 0.03408    & 0.02266     & 0.01962     & 0.1094    & 0.05296  \\
1       & 4.2806  & 0.1126    & 0.006002  & 0.2547     & 0.08623     & 0.001284    & 0.2274    & 0.02112  \\
1       & 5.0092  & 0.1099    & 0.05391   & 0.005768   & 0.000007733 & 0.02135     & 0.05215   & 0.001514 \\
1       & 5.4438  & 0.2641    & 0.01765   & 0.002      & 0.0407      & 0.04815     & 0.01973   & 0.07193  \\
1       & 5.5356  & 0.01798   & 0.1644    & 0.00008815 & 0.0521      & 0.000311    & 0.05749   & 0.1974   \\
1       & 5.8152  & 0.007681  & 0.0009797 & 0.08981    & 0.005305    & 0.02411     & 0.01974   & 0.04288  \\
1       & 6.1044  & 0.1267    & 0.05321   & 0.05725    & 0.1904      & 0.008151    & 0.0002141 & 0.04928  \\
1       & 6.377   & 0.004254  & 0.1858    & 0.01022    & 0.04235     & 0.009649    & 0.07419   & 0.01484  \\
1       & 6.53    & 0.1333    & 0.005649  & 0.178      & 0.02216     & 0.01075     & 0.001722  & 0.1219   \\
1       & 6.6492  & 4.442E-07 & 0.002534  & 0.02368    & 0.02788     & 0.02103     & 0.008559  & 0.01298  \\
1       & 6.9034  & 0.0108    & 0.06437   & 0.007711   & 0.002004    & 0.00331     & 0.2131    & 0.000833 \\
1       & 7.0252  & 0.004178  & 0.01263   & 0.001188   & 0.006141    & 0.005816    & 0.007048  & 0.1654   \\
1       & 7.1253  & 0.01661   & 0.01223   & 0.0693     & 0.009195    & 0.06725     & 0.02139   & 0.09644  \\
1       & 7.4944  & 0.001409  & 0.00322   & 0.01293    & 0.1612      & 0.000005525 & 0.164     & 0.04731  \\
1       & 7.5839  & 0.004254  & 0.0247    & 0.1047     & 0.0002221   & 0.01114     & 0.1066    & 0.000118 \\
1       & 7.6748  & 0.0901    & 0.1146    & 0.001978   & 0.1056      & 0.01438     & 0.1473    & 0.1429   \\
1       & 7.8701  & 0.03842   & 0.0001167 & 0.007617   & 0.00142     & 0.002646    & 0.004836  & 0.1535   \\
1       & 7.9942  & 0.2083    & 0.0004034 & 0.00796    & 0.1915      & 0.01358     & 0.007824  & 0.1374  \\
\hline 
\end{tabular}
\end{table}

\clearpage{}\newpage{}

\begin{table}
\global\long\def\thetable{XXIV}
 \centering \caption{$^{46}$Ti J=1 (columns) to J=0 (rows) - Orbit}
\begin{tabular}{c | c | c c c c c c c}
\hline 
        & Isospin & 1          & 1         & 1          & 1           & 1          & 1         & 1         \\
\hline
Isopsin & Eshift  & 3.9959     & 5.6008    & 6.0668     & 6.6919      & 7.0243     & 7.1855    & 7.4875    \\
\hline
1       & 0       & 0.1149     & 0.03794   & 0.01696    & 0.00000284  & 0.005989   & 0.008404  & 0.007926  \\
1       & 4.2488  & 0.04915    & 0.0001792 & 0.1067     & 3.491E-07   & 0.009186   & 0.04022   & 0.01652   \\
1       & 5.2874  & 0.2494     & 0.0382    & 0.001331   & 0.000007078 & 0.005935   & 0.001301  & 0.0249    \\
1       & 5.6685  & 0.01483    & 0.04599   & 0.00003105 & 0.08303     & 0.0744     & 0.0129    & 0.00561   \\
1       & 6.58    & 0.01038    & 0.03509   & 0.1984     & 0.08515     & 0.04006    & 0.01695   & 0.005311  \\
1       & 6.8885  & 0.003817   & 0.001405  & 0.005974   & 0.00009793  & 0.01656    & 0.02226   & 0.0003261 \\
1       & 7.3222  & 0.02747    & 0.02007   & 0.01101    & 0.0006847   & 0.002182   & 0.009649  & 0.003563  \\
1       & 8.1666  & 0.09741    & 0.002786  & 0.06559    & 0.002193    & 0.02062    & 0.05959   & 0.03684   \\
1       & 8.5605  & 0.02953    & 0.002211  & 0.03216    & 0.03094     & 0.04768    & 0.000796  & 0.007777  \\
1       & 9.0534  & 0.005026   & 0.008423  & 0.00003215 & 0.0001571   & 0.003314   & 0.00147   & 0.0214    \\
1       & 9.4385  & 0.001808   & 0.01665   & 0.002379   & 0.02719     & 0.001692   & 0.0009588 & 0.01853   \\
1       & 9.9011  & 0.0004115  & 0.006736  & 0.00005477 & 0.03064     & 0.03135    & 0.01678   & 0.01289   \\
1       & 10.0432 & 0.00002369 & 0.0003114 & 0.002971   & 0.0009833   & 0.00003019 & 0.00245   & 0.0004052 \\
1       & 10.1636 & 0.0001251  & 0.005331  & 0.000448   & 0.01571     & 0.002439   & 0.02292   & 0.002768  \\
1       & 10.3307 & 0.004025   & 0.009075  & 0.00004459 & 0.0006423   & 0.005499   & 0.001417  & 0.005892  \\
1       & 10.6885 & 0.004425   & 0.002795  & 0.000027   & 0.0002572   & 0.001279   & 0.0002074 & 0.0008467 \\
1       & 10.8004 & 0.00374    & 0.002765  & 0.00005227 & 0.005215    & 0.00001634 & 0.001119  & 0.006717  \\
1       & 11.0128 & 0.0002405  & 0.002663  & 0.006632   & 0.002503    & 0.00007545 & 0.000208  & 0.001167  \\
1       & 11.2704 & 0.0002503  & 0.004969  & 0.00004405 & 0.000001502 & 0.00021    & 0.01228   & 0.0004441 \\
1       & 11.4117 & 0.0001154  & 0.00272   & 0.002269   & 0.002497    & 0.00002457 & 0.0006746 & 0.001893 \\
\hline 
\end{tabular}

\bigskip

\global\long\def\thetable{XXV}
 \centering \caption{$^{46}$Ti J=1 (columns) to J=2 (rows) - Orbit}
\begin{tabular}{c | c | c c c c c c c}
\hline 
        & Isospin & 1         & 1         & 1           & 1          & 1           & 1          & 1         \\
\hline
Isopsin & Eshift  & 3.9959    & 5.6008    & 6.0668      & 6.6919     & 7.0243      & 7.1855     & 7.4875    \\
\hline
1       & 1.0055  & 0.03343   & 0.001889  & 0.06257     & 0.00007682 & 0.01682     & 0.001936   & 0.007752  \\
1       & 2.5869  & 0.2048    & 0.0566    & 0.0002216   & 0.003332   & 0.03743     & 0.00249    & 0.001937  \\
1       & 3.3857  & 0.0002012 & 0.02181   & 0.008233    & 0.007793   & 0.00849     & 0.01038    & 0.01386   \\
1       & 4.2806  & 0.09203   & 0.003008  & 0.1398      & 0.06733    & 0.008613    & 0.02059    & 0.006669  \\
1       & 5.0092  & 0.0004329 & 0.005207  & 0.005446    & 0.01083    & 0.01333     & 0.01258    & 0.01467   \\
1       & 5.4438  & 0.05836   & 0.001339  & 0.001775    & 0.0221     & 0.03091     & 0.03043    & 0.0008939 \\
1       & 5.5356  & 0.03535   & 0.009421  & 0.001918    & 0.000812   & 0.01135     & 0.01163    & 0.02744   \\
1       & 5.8152  & 0.06813   & 0.007087  & 0.1057      & 0.0003126  & 0.0321      & 0.0003941  & 0.0003432 \\
1       & 6.1044  & 0.02153   & 0.01682   & 0.08962     & 0.06327    & 0.00008514  & 0.001561   & 0.001057  \\
1       & 6.377   & 0.0006567 & 0.07386   & 0.02197     & 0.005356   & 0.007608    & 0.00002165 & 0.0001444 \\
1       & 6.53    & 0.02139   & 0.00115   & 0.02406     & 0.003882   & 0.009383    & 0.02047    & 0.0007482 \\
1       & 6.6492  & 0.02777   & 0.0001458 & 0.002423    & 0.002815   & 0.000001403 & 0.002737   & 0.001856  \\
1       & 6.9034  & 0.004958  & 0.005625  & 0.01315     & 0.02495    & 0.001183    & 0.02011    & 0.01485   \\
1       & 7.0252  & 0.0002675 & 0.0002264 & 0.005438    & 0.001626   & 0.04455     & 0.000522   & 0.0001361 \\
1       & 7.1253  & 0.02937   & 0.01074   & 0.000003221 & 0.01524    & 0.0504      & 0.003563   & 0.01442   \\
1       & 7.4944  & 0.001082  & 0.0001034 & 0.002792    & 0.02661    & 0.01207     & 0.02492    & 0.03046   \\
1       & 7.5839  & 0.01295   & 0.02734   & 0.04101     & 0.003033   & 0.01248     & 0.04723    & 0.002822  \\
1       & 7.6748  & 0.001532  & 0.0006948 & 0.00196     & 0.05307    & 0.0007214   & 0.02328    & 0.01063   \\
1       & 7.8701  & 0.008382  & 0.01743   & 0.004208    & 0.0008357  & 0.01249     & 0.04336    & 0.02686   \\
1       & 7.9942  & 0.0305    & 0.002178  & 0.000106    & 0.009564   & 0.01313     & 0.004865   & 0.004333 \\
\hline 
\end{tabular}
\end{table}

\clearpage{}\newpage{}

\begin{table}[h]
\global\long\def\thetable{XXVI}
 \centering \caption{$^{44}$Ti J=1 (columns) to J=0 (rows) - Total}
\begin{tabular}{c | c | c c c c c c c}
\hline 
        & Isospin & 1          & 1          & 0         & 1          & 1           & 0           & 1           \\
\hline
Isopsin & Eshift  & 5.965      & 7.9716     & 8.8348    & 9.3734     & 9.5168      & 9.5715      & 10.1337     \\
\hline
        & Isospin & 1          & 1          & 0         & 1          & 1           & 0           & 1           \\
Isopsin & Eshift  & 5.965      & 7.9716     & 8.8348    & 9.3734     & 9.5168      & 9.5715      & 10.1337     \\
0       & 0       & 1.503      & 0.1808     & 0.009623  & 0.01502    & 0.00002501  & 0.000004861 & 0.02056     \\
0       & 4.5009  & 0.2049     & 0.5646     & 0.001427  & 0.4162     & 0.006138    & 0.0002452   & 0.01264     \\
0       & 7.6338  & 0.09797    & 1.013      & 0.001574  & 0.5605     & 0.7152      & 0.0001405   & 0.2339      \\
0       & 8.2697  & 0.02169    & 3.328      & 0.0005478 & 0.04633    & 1.86        & 0.0003522   & 0.2196      \\
2       & 8.8928  & 2.675      & 1.827      & 0         & 1.888      & 0.5141      & 3.751E-09   & 0.002571    \\
0       & 9.2269  & 0.4547     & 0.7503     & 0.002911  & 0.006015   & 1.001       & 0.000004064 & 0.06647     \\
0       & 10.1805 & 0.2331     & 0.2192     & 0.001373  & 1.8        & 0.6513      & 0.00009808  & 0.06615     \\
1       & 10.2814 & 0.0001105  & 0.0005071  & 0.2351    & 0.00007183 & 0.000002939 & 0.0008016   & 0.000251    \\
0       & 11.1362 & 0.006506   & 0.1005     & 0.000457  & 0.002859   & 0.3957      & 0.00002896  & 0.2174      \\
1       & 11.5847 & 0.002657   & 0.006127   & 0.09646   & 0.00002522 & 0.0001197   & 0.04115     & 0.0006762   \\
0       & 12.1278 & 0.007523   & 0.03866    & 0.0004182 & 0.07007    & 0.0002538   & 0.00007129  & 0.09656     \\
1       & 12.8973 & 0.0006278  & 0.00004375 & 0.1248    & 0.0001264  & 0.00005995  & 0.0008475   & 0.001015    \\
0       & 13.1724 & 0.007962   & 0.03473    & 0.001441  & 0.03995    & 0.1539      & 0.00005685  & 0.0804      \\
1       & 13.4225 & 0.0008335  & 0.00004586 & 0.4221    & 0.0008047  & 0.00001428  & 0.1094      & 0.0004633   \\
0       & 13.7097 & 0.02171    & 0.2097     & 0.004708  & 0.456      & 0.05265     & 0.0004453   & 0.01595     \\
1       & 13.9211 & 0.0002448  & 0.0006254  & 0.1966    & 0.0002995  & 0.0001577   & 0.02198     & 0.000007402 \\
2       & 13.971  & 0.03042    & 0.00259    & 7.723E-09 & 0.1322     & 0.001938    & 3.105E-09   & 0.2674      \\
1       & 14.1648 & 0.003909   & 0.0002751  & 1.192     & 0.00004685 & 0.00001816  & 0.02586     & 0.000422    \\
0       & 14.2332 & 0.002762   & 0.00265    & 0.0007879 & 0.0004155  & 0.003352    & 0.0006678   & 0.1728      \\
1       & 15.3154 & 0.00002524 & 0.0000105  & 0.0001772 & 0.00002168 & 0.000000979 & 0.004588    & 0.0003454  \\
\hline 
\end{tabular}

\bigskip

\global\long\def\thetable{XXVII}
 \centering \caption{$^{44}$Ti J=1 (columns) to J=2 (rows) - Total}
\begin{tabular}{c | c | c c c c c c c}
\hline 
        & Isospin & 1         & 1          & 0           & 1          & 1           & 0          & 1          \\
\hline
Isopsin & Eshift  & 5.965     & 7.9716     & 8.8348      & 9.3734     & 9.5168      & 9.5715     & 10.1337    \\
\hline
0       & 1.2874  & 0.1209    & 0.3976     & 0.004955    & 0.4648     & 0.07475     & 0.0001428  & 0.15       \\
0       & 3.1681  & 2.031     & 0.01111    & 0.006597    & 0.005543   & 0.09305     & 0.00005383 & 0.2223     \\
1       & 5.2972  & 0.00382   & 0.001909   & 1.949       & 0.0000319  & 0.0001648   & 0.1128     & 0.00005778 \\
0       & 6.4926  & 0.293     & 0.1599     & 0.000006072 & 0.003145   & 0.5926      & 0.00005244 & 0.2434     \\
0       & 6.768   & 5.512     & 3.693      & 0.004765    & 2.17       & 0.761       & 0.00002982 & 0.1441     \\
0       & 7.2394  & 0.09388   & 1.714      & 0.002189    & 0.03073    & 0.003586    & 0.00009255 & 1.05       \\
1       & 7.8878  & 0.0001591 & 0.0005239  & 0.005455    & 0.0007804  & 0.0001831   & 0.3635     & 0.001406   \\
0       & 8.5457  & 0.0006127 & 0.6773     & 0.0005068   & 0.9249     & 3.163       & 2.661E-07  & 0.8189     \\
0       & 8.8657  & 0.001142  & 0.1978     & 0.00005242  & 0.1709     & 0.01596     & 0.00000576 & 0.1711     \\
1       & 8.947   & 0.0007049 & 0.00005597 & 0.2779      & 0.00009404 & 0.00008692  & 0.1882     & 0.001511   \\
1       & 9.3489  & 0.0003826 & 0.001331   & 0.3678      & 0.002134   & 0.000004114 & 0.6805     & 0.00007917 \\
0       & 9.4628  & 0.002154  & 0.2373     & 0.001424    & 0.9494     & 0.1147      & 0.001665   & 0.5832     \\
1       & 9.4677  & 0.001027  & 0.0006549  & 0.3442      & 0.00002526 & 0.0008856   & 0.2489     & 0.0003305  \\
0       & 9.6586  & 0.1932    & 0.6258     & 0.001062    & 0.3561     & 0.5635      & 0.00002798 & 2.619      \\
1       & 9.7447  & 0.001816  & 0.00001226 & 0.4057      & 0.0001849  & 0.00126     & 0.08427    & 0.0003943  \\
0       & 9.8125  & 0.04605   & 1.193      & 0.00007475  & 0.4343     & 1.343       & 0.00005085 & 0.005515   \\
2       & 10.2546 & 0.003072  & 3.33       & 7.169E-12   & 0.05114    & 0.9437      & 1.944E-08  & 0.1189     \\
0       & 10.2872 & 0.008542  & 0.03862    & 0.0001683   & 0.08189    & 0.0659      & 0.0001638  & 0.02096    \\
1       & 10.6656 & 0.0001629 & 0.000158   & 0.002461    & 0.0004166  & 0.001121    & 0.06816    & 0.002358   \\
0       & 10.7354 & 0.4132    & 1.115      & 0.004939    & 2.622      & 1.794       & 0.0002026  & 0.6642    \\
\hline 
\end{tabular}
\end{table}

\clearpage{}\newpage{}

\begin{table}[h]
\global\long\def\thetable{XXVIII}
 \centering \caption{$^{44}$Ti J=1 (columns) to J=0 (rows) - Spin}
\begin{tabular}{c | c | c c c c c c c}
\hline 
        & Isospin & 1         & 1          & 0           & 1          & 1           & 0          & 1          \\
\hline
Isopsin & Eshift  & 5.965     & 7.9716     & 8.8348      & 9.3734     & 9.5168      & 9.5715     & 10.1337    \\
\hline
0       & 0       & 0.6249    & 0.03334    & 0.05162     & 0.1181     & 0.01384     & 0.00002625 & 0.07818    \\
0       & 4.5009  & 0.1882    & 0.1267     & 0.007655    & 0.09783    & 0.002912    & 0.001315   & 0.006332   \\
0       & 7.6338  & 0.1345    & 0.3375     & 0.008455    & 0.1119     & 0.1423      & 0.0007456  & 0.06906    \\
0       & 8.2697  & 0.02618   & 1.421      & 0.002943    & 0.08675    & 0.5612      & 0.001909   & 0.1115     \\
2       & 8.8928  & 1.036     & 0.538      & 0           & 1.188      & 0.4263      & 4.418E-09  & 0.02974    \\
0       & 9.2269  & 0.454     & 0.2266     & 0.01559     & 0.01439    & 0.1941      & 0.00002364 & 0.0009815  \\
0       & 10.1805 & 0.3921    & 0.2229     & 0.007372    & 1.004      & 0.2775      & 0.0005182  & 0.03029    \\
1       & 10.2814 & 0.0005913 & 0.002723   & 0.2309      & 0.0003846  & 0.00001548  & 0.0008962  & 0.001348   \\
0       & 11.1362 & 0.001263  & 0.03893    & 0.002444    & 0.00004475 & 0.1422      & 0.0001533  & 0.2306     \\
1       & 11.5847 & 0.01425   & 0.03287    & 0.08394     & 0.0001351  & 0.0006426   & 0.01834    & 0.003619   \\
0       & 12.1278 & 0.008678  & 0.01271    & 0.002246    & 0.08918    & 0.00651     & 0.0003819  & 0.1676     \\
1       & 12.8973 & 0.003368  & 0.0002352  & 0.06433     & 0.0006798  & 0.0003211   & 0.008023   & 0.00545    \\
0       & 13.1724 & 0.01856   & 0.01753    & 0.007724    & 0.01175    & 0.07322     & 0.0003015  & 0.01605    \\
1       & 13.4225 & 0.004473  & 0.000245   & 0.3315      & 0.004317   & 0.0000774   & 0.01579    & 0.002488   \\
0       & 13.7097 & 0.01239   & 0.1105     & 0.02523     & 0.2427     & 0.0307      & 0.002387   & 0.00005815 \\
1       & 13.9211 & 0.001311  & 0.003351   & 0.2009      & 0.00162    & 0.0008412   & 0.01055    & 0.00003875 \\
2       & 13.971  & 0.01195   & 0.0003502  & 9.668E-09   & 0.03311    & 5.825E-07   & 1.436E-09  & 0.1365     \\
1       & 14.1648 & 0.02095   & 0.001477   & 0.8147      & 0.0002534  & 0.00009551  & 0.00262    & 0.002265   \\
0       & 14.2332 & 0.005269  & 0.007607   & 0.004227    & 0.002056   & 0.00197     & 0.003576   & 0.1056     \\
1       & 15.3154 & 0.0001345 & 0.00005609 & 0.000002275 & 0.0001167  & 0.000005393 & 0.01607    & 0.001859  \\
\hline 
\end{tabular}

\bigskip

\global\long\def\thetable{XXIX}
 \centering \caption{$^{44}$Ti J=1 (columns) to J=2 (rows) - Spin}
\begin{tabular}{c | c | c c c c c c c}
\hline 
        & Isospin & 1         & 1          & 0          & 1         & 1          & 0          & 1         \\
\hline
Isopsin & Eshift  & 5.965     & 7.9716     & 8.8348     & 9.3734    & 9.5168     & 9.5715     & 10.1337   \\
\hline
0       & 1.2874  & 0.2441    & 0.05143    & 0.02658    & 0.3393    & 0.04309    & 0.0007635  & 0.08232   \\
0       & 3.1681  & 0.8927    & 0.01685    & 0.03538    & 0.005598  & 0.0174     & 0.0002902  & 0.1559    \\
1       & 5.2972  & 0.02047   & 0.01023    & 1.796      & 0.0001708 & 0.0008918  & 0.0271     & 0.0003078 \\
0       & 6.4926  & 0.04836   & 0.01052    & 0.00003255 & 0.04767   & 0.2189     & 0.0002845  & 0.09342   \\
0       & 6.768   & 2.849     & 1.052      & 0.02559    & 1.099     & 0.3688     & 0.0001654  & 0.02648   \\
0       & 7.2394  & 0.003497  & 0.6712     & 0.01173    & 0.01187   & 0.0009394  & 0.0004962  & 0.4351    \\
1       & 7.8878  & 0.0008539 & 0.002802   & 0.003638   & 0.004183  & 0.0009756  & 0.1049     & 0.007553  \\
0       & 8.5457  & 0.002541  & 0.3105     & 0.002729   & 0.6819    & 1.16       & 0.00000228 & 0.554     \\
0       & 8.8657  & 0.0002612 & 0.1713     & 0.0002849  & 0.1566    & 0.01669    & 0.00002893 & 0.1047    \\
1       & 8.947   & 0.003777  & 0.0003018  & 0.1816     & 0.0005055 & 0.0004709  & 0.1262     & 0.00811   \\
1       & 9.3489  & 0.002054  & 0.007142   & 0.1385     & 0.01145   & 0.00002431 & 0.2038     & 0.0004272 \\
0       & 9.4628  & 0.0003618 & 0.1001     & 0.007659   & 0.4372    & 0.09155    & 0.008898   & 0.4737    \\
1       & 9.4677  & 0.005509  & 0.00353    & 0.1661     & 0.0001314 & 0.004767   & 0.03992    & 0.001787  \\
0       & 9.6586  & 0.03921   & 0.04563    & 0.00569    & 0.1595    & 0.2579     & 0.000152   & 1.66      \\
1       & 9.7447  & 0.009742  & 0.00006594 & 0.1462     & 0.0009895 & 0.006751   & 0.05149    & 0.002118  \\
0       & 9.8125  & 0.02104   & 0.4069     & 0.0003971  & 0.32      & 0.7295     & 0.0002808  & 0.08963   \\
2       & 10.2546 & 0.0328    & 0.9406     & 7.169E-12  & 0.0002878 & 0.2919     & 7.251E-09  & 0.1547    \\
0       & 10.2872 & 0.05891   & 0.009018   & 0.0009003  & 0.1327    & 0.0346     & 0.0008786  & 0.002554  \\
1       & 10.6656 & 0.0008694 & 0.0008448  & 0.001203   & 0.002249  & 0.006007   & 0.06752    & 0.01265   \\
0       & 10.7354 & 0.2547    & 0.8166     & 0.02646    & 1.059     & 0.9423     & 0.001077   & 0.2657   \\
\hline 
\end{tabular}
\end{table}

\clearpage{}\newpage{}

\begin{table}[h]
\global\long\def\thetable{XXX}
 \centering \caption{$^{44}$Ti J=1 (columns) to J=0 (rows) - Orbit}
\begin{tabular}{c | c | c c c c c c c}
\hline 
        & Isospin & 1          & 1          & 0          & 1          & 1           & 0           & 1          \\
\hline
Isopsin & Eshift  & 5.965      & 7.9716     & 8.8348     & 9.3734     & 9.5168      & 9.5715      & 10.1337    \\
\hline
0       & 0       & 0.1897     & 0.05888    & 0.01666    & 0.04889    & 0.01268     & 0.000008519 & 0.01856    \\
0       & 4.5009  & 0.0003531  & 0.1563     & 0.002471   & 0.1105     & 0.0005943   & 0.0004247   & 0.00108    \\
0       & 7.6338  & 0.002882   & 0.1812     & 0.002733   & 0.1715     & 0.2194      & 0.0002387   & 0.04876    \\
0       & 8.2697  & 0.0002105  & 0.3999     & 0.0009511  & 0.006286   & 0.3779      & 0.000621    & 0.01813    \\
2       & 8.8928  & 0.3819     & 0.3822     & 0          & 0.0807     & 0.004106    & 2.728E-11   & 0.01482    \\
0       & 9.2269  & 2.785E-07  & 0.1522     & 0.005026   & 0.001799   & 0.3136      & 0.000008103 & 0.05129    \\
0       & 10.1805 & 0.02058    & 0.00001554 & 0.002382   & 0.1153     & 0.07856     & 0.0001654   & 0.006916   \\
1       & 10.2814 & 0.0001906  & 0.00088    & 0.00001876 & 0.000124   & 0.000004927 & 0.00000264  & 0.0004358  \\
0       & 11.1362 & 0.002036   & 0.01433    & 0.0007873  & 0.002189   & 0.06346     & 0.00004898  & 0.0001954  \\
1       & 11.5847 & 0.004601   & 0.01061    & 0.0004351  & 0.0000436  & 0.0002076   & 0.004549    & 0.001166   \\
0       & 12.1278 & 0.00004118 & 0.007036   & 0.0007257  & 0.00115    & 0.004193    & 0.0001232   & 0.009729   \\
1       & 12.8973 & 0.001087   & 0.00007605 & 0.009916   & 0.0002199  & 0.0001035   & 0.01409     & 0.001761   \\
0       & 13.1724 & 0.002208   & 0.002909   & 0.002492   & 0.008371   & 0.01483     & 0.00009649  & 0.0246     \\
1       & 13.4225 & 0.001445   & 0.00007888 & 0.005469   & 0.001394   & 0.00002519  & 0.04206     & 0.0008041  \\
0       & 13.7097 & 0.0013     & 0.01577    & 0.008138   & 0.03335    & 0.002941    & 0.0007701   & 0.01794    \\
1       & 13.9211 & 0.0004228  & 0.001081   & 0.00002312 & 0.0005264  & 0.0002705   & 0.002075    & 0.00001228 \\
2       & 13.971  & 0.004238   & 0.004844   & 1.091E-10  & 0.03297    & 0.001871    & 3.177E-10   & 0.02177    \\
1       & 14.1648 & 0.006759   & 0.0004773  & 0.03577    & 0.00008236 & 0.00003037  & 0.01202     & 0.000732   \\
0       & 14.2332 & 0.0004013  & 0.001277   & 0.001365   & 0.0006229  & 0.0001827   & 0.001153    & 0.008234   \\
1       & 15.3154 & 0.00004318 & 0.00001805 & 0.0002197  & 0.00003777 & 0.000001777 & 0.003487    & 0.0006018 \\
\hline 
\end{tabular}

\bigskip

\global\long\def\thetable{XXXI}
 \centering \caption{$^{44}$Ti J=1 (columns) to J=2 (rows) - Orbit}
\begin{tabular}{c | c | c c c c c c c}
\hline 
        & Isospin & 1         & 1          & 0          & 1          & 1           & 0           & 1          \\
\hline
Isopsin & Eshift  & 5.965     & 7.9716     & 8.8348     & 9.3734     & 9.5168      & 9.5715      & 10.1337    \\
\hline
0       & 1.2874  & 0.02141   & 0.163      & 0.008582   & 0.009848   & 0.004333    & 0.0002459   & 0.01009    \\
0       & 3.1681  & 0.2307    & 0.0005949  & 0.01142    & 0.02228    & 0.02997     & 0.00009407  & 0.005875   \\
1       & 5.2972  & 0.006605  & 0.003299   & 0.003125   & 0.00005507 & 0.0002899   & 0.02934     & 0.00009886 \\
0       & 6.4926  & 0.1033    & 0.08837    & 0.0000105  & 0.02633    & 0.09115     & 0.00009268  & 0.03523    \\
0       & 6.768   & 0.4354    & 0.8033     & 0.008269   & 0.1804     & 0.07023     & 0.00005479  & 0.04702    \\
0       & 7.2394  & 0.06114   & 0.2401     & 0.003786   & 0.004403   & 0.0008544   & 0.0001602   & 0.1333     \\
1       & 7.8878  & 0.0002759 & 0.0009026  & 0.0001835  & 0.00135    & 0.0003134   & 0.07786     & 0.002441   \\
0       & 8.5457  & 0.005648  & 0.07064    & 0.0008839  & 0.01849    & 0.4919      & 9.881E-07   & 0.02579    \\
0       & 8.8657  & 0.000311  & 0.0009579  & 0.00009289 & 0.0003123  & 0.000008332 & 0.000008874 & 0.00811    \\
1       & 8.947   & 0.001219  & 0.00009783 & 0.01021    & 0.0001635  & 0.0001532   & 0.006173    & 0.002619   \\
1       & 9.3489  & 0.0006637 & 0.002307   & 0.05489    & 0.003697   & 0.00000842  & 0.1395      & 0.0001386  \\
0       & 9.4628  & 0.0007504 & 0.02912    & 0.002478   & 0.09806    & 0.001301    & 0.002864    & 0.005681   \\
1       & 9.4677  & 0.001779  & 0.001144   & 0.03208    & 0.00004143 & 0.001543    & 0.08949     & 0.0005803  \\
0       & 9.6586  & 0.05831   & 0.3335     & 0.001836   & 0.03896    & 0.05899     & 0.00004954  & 0.1088     \\
1       & 9.7447  & 0.003145  & 0.00002134 & 0.06481    & 0.000319   & 0.002177    & 0.004015    & 0.0006845  \\
0       & 9.8125  & 0.1293    & 0.2063     & 0.0001273  & 0.008713   & 0.09283     & 0.00009267  & 0.05068    \\
2       & 10.2546 & 0.01579   & 0.7309     & 4.897E-33  & 0.0591     & 0.1859      & 2.947E-09   & 0.002354   \\
0       & 10.2872 & 0.02258   & 0.01031    & 0.0002901  & 0.006107   & 0.005       & 0.0002836   & 0.00888    \\
1       & 10.6656 & 0.0002797 & 0.0002722  & 0.0002229  & 0.0007295  & 0.001938    & 0.000001512 & 0.004082   \\
0       & 10.7354 & 0.01909   & 0.02314    & 0.008534   & 0.348      & 0.136       & 0.0003455   & 0.08971   \\
\hline 
\end{tabular}
\end{table}

\restoregeometry

\end{document}